\documentclass[12pt]{article}
\usepackage{scicite}
\usepackage{epsfig}

\newenvironment{sciabstract}{%
\begin{quote} \bf}
{\end{quote}}

\title{Experimental demonstration of counterfactual quantum communication}

\author
{Yang Liu$^\dagger$, Lei Ju$^\dagger$,  Xiao-Lei Liang, \\
\\
Shi-Biao Tang, Guo-Liang Shen Tu, Lei Zhou, Cheng-Zhi Peng, \\
\\
Kai Chen, Teng-Yun Chen$^\ast$, Zeng-Bing Chen,and Jian-Wei Pan$^\ast$ \\
\\
\normalsize{Hefei National Laboratory for Physical Sciences at Microscale}\\
\normalsize{ and Department of Modern Physics,}\\
\normalsize{University of Science and Technology of China, Hefei, Anhui 230026, China}\\
\\
\normalsize{$^\dagger$These authors contributed equally to this work.}\\
\normalsize{$^\ast$To whom correspondence should be addressed. E-mail: tychen@ustc.edu.cn,pan@ustc.edu.cn}
}

\date{}

\begin{document}

\baselineskip24pt

\maketitle

\begin{sciabstract}
Based on principle of quantum mechanics, quantum cryptography
provides an intriguing way to establish secret keys between
remote parties, generally relying on actual transmission of signal
particles. Surprisingly, an even more striking method is recently proposed by
Noh \cite{Noh2009} named as `counterfactual quantum cryptography'
enabling key distribution, in which particles carrying secret
information are seemly not being transmitted through quantum
channel. We experimentally give here a faithful implementation by
following the scheme with an on-table realization. Furthermore, we
report an illustration on a 1 km fiber operating at telecom
wavelength to verify its feasibility for extending to long
distance. For both cases, high visibilities of better than 98\% are
maintained with active stabilization of interferometers, while a quantum bit
error rate around 5.5\% is attained after 1 km channel.
\end{sciabstract}

The principle of quantum mechanics provides exciting opportunities
across a number of applications which offer high-performance and
superiority over classical ways, among which quantum key
distribution appears as the first practical application with
provable unconditional security between communication parties. Since
proposal of BB84 protocol\cite{BB84} for quantum key distribution, there
have spurred a flurry of activity in both theoretical developments
and successful experimental demonstrations of quantum cryptography
systems ranging from a two-party communication up to a network
\cite{BB84,Lo-Chau1999,Shor-Preskill,Gisin2002,Hwang2003,GLLP,Wang2005,Lo-Ma-Chen2005,Elliott2005,ILM,Pan-decoy,America-decoy,EU-decoy,Toshiba-decoy,SECOQC,review2009,Qiang09,Guo2009,Chen2010,TokyoQKD}.
Particularly, recent advances have covered
high speed and long distance demonstrations under imperfect devices,
by utilizing decoy state method
\cite{Hwang2003,Wang2005,Lo-Ma-Chen2005,Pan-decoy,America-decoy,EU-decoy,Toshiba-decoy,SECOQC,Guo2009,Chen2010,TokyoQKD}.
In order to produce identical secure keys between two remote communication parties,
key distribution process generally require actual transmission of signal particles through a quantum channel,
before detections and data post-processing.

In conformity with quantum mechanics, there are many fundamental quantum phenomena
that lead to intriguing physical effects
\cite{ElitzurVaidman1993,Kwiat1999,Mitchison-Jozsa2001,kwiat2006,noncommuting2007}.
Striking applications are consequently proposed like interaction-free measurement
\cite{ElitzurVaidman1993}, counterfactual quantum computation
\cite{Mitchison-Jozsa2001}, in which the presence of a nontransmitting object is
ascertained seemingly without interacting with it, or outcomes of computations can
sometimes be inferred without running of a computer, as successfully demonstrated
in \cite{Kwiat1999} and \cite{kwiat2006}, respectively. Following this route, Noh
has presented an engaging method to do key distribution, named as `counterfactual
quantum cryptography' \cite{Noh2009}. This engrossing proposal gives not only a
conceptually new approach to do secure communication by use of counterfactual
characterization in quantum phenomena, but also it promises potential security
advantage for communication. In the scheme, quantum signals producing secure final
keys are not actually transmitted. Recently an unconditional security proof is
provided in the case of ideal case by utilizing perfect single-photon source and
with well-controlled detectors' efficiencies etc \cite{Guo2010}. An attempt is
made in \cite{Zeng2011}, in which a modified realization of the proposal is
presented. The original scheme is, however, not be justified by the demonstration,
whose equivalence and security need additionally thorough analysis. Fascinatingly,
the counterfactual method itself provides an employment of quantum principles for
applicable task, and motivates us a more novel conceptually thinking for
fundamental quantum mechanics.

We present here experimentally a faithful implementation of the counterfactual
quantum communication (CQC) operating at telecom
wavelength. we have first set up the experiment on a table, and
consequently extended it to a 1 km fiber to check feasibility for
long-distance realization. Active feed-back control technology is
developed for real-time stabilization of Michelson-type
interferometer utilized in the scheme, to guarantee favorable
condition for successful operation of quantum communication. In
addition, tailored optical and controlling electronics designs are
made, which enables to attain a high visibility of better than 98\%
for interferometer through 1 km long fiber.

A schematic of the experimental setup is illustrated in Fig. 1. Two distributed
feedback laser diodes (LDs) with pulse width about 1 ns for
generating horizontal and vertical polarizations are combined with a
polarization beam splitter (PBS) to act as signal light. The center
wavelengths of the two LDs were carefully tuned to be consistent
within a range of 0.02 nm. Random number will be supplied to enable
Alice to generate a sequence of pulses with random polarization
$|H>$ or $|V>$. An attenuator is used to attenuate the pulse to
single photon level at Alice's output port. As depicted by the
protocol, the signal pulse is split by a beam splitter (BS), and
then travels through arm $a$ or arm $b$ for transmission. A
polarization controller (PC) is inserted for compensating possible
polarization drift of fiber channel. In the protocol, Bob will
randomly choose one of the two polarizations representing his bit
value. The optical path $b$ of the single-photon pulse will be
blocked if the polarization of the pulse is identical to his
polarization. For achieving this aim, a tailored PBS is used with
the reflecting port connected with a jumper, which is about 20 m
long, enough for accumulating time needed for changing status of
optical switch (SW). This design allows to set a delay for H and V
pulse: horizontal light will pass the PBS directly, while vertical
light will reflect to the delay line, and then transmit back to
another output port of PBS before going through SW. Bob can
therefore control accurately switch timing to guide signal photon to
be directly detected by his polarization sensitive detector, or to
be rotated and reflected by Faraday mirror (FM) and to return back
to BS. In our experiment, polarization sensitive detecting is
accomplished by a PBS and two single photon detectors. Therefore
through accurate control of the switch timing, Bob can determine
effectively whether switch the polarization state to the detector
$D3$ or not.

If Bob chooses a different polarization from Alice, the pulse will be reflected by Bob and combined again at BS. In this
case, the photonic pulse will go toward detector $D2$ due to interference effect if there is a phase difference of $\pi$
radians between the two paths $a$ and $b$.  An attenuator is added to arm $a$ to balance the optical loss
between two arms. To compensate the phase drift in the fiber, and to be sure that all light after interference will
go to detector $D2$, a fast fiber stretcher (FS) is used for compensating fast phase variation in the two arms.
Moreover, an optical delay line (OD) is employed by Alice to adjust the path difference between arm $a$ and arm $b$ of
the whole system with an precision of less than 0.1mm. In order to obtain good visibility in the interferometer,
feedback control is further developed by using a laser with different wavelength of 1570 nm and controlling electronics.
The feedback laser and the signal laser is combined by a dense wavelength division multiplexer (DWDM) before
entering BS.

If their choices are same, the interferometer will be destroyed,
since Bob chooses to detect rather than let it go through, either
$D1,D2$ of Alice or $D3$ of Bob will detect a pulse signal. We save
only events for $D1$ clicks, and kept all other events as a
monitoring of Eve's intervention. For $D1$'s clicks, only the events
for which $D1$ detects a correct final polarization will be kept by
Alice and no information is revealed, otherwise measurements results
are announced. For $D2$ or $D3$'s clicks, both detected polarization
and initial polarization states are released for detection of
possible eavesdropping. Accord to the scheme, the initial quantum
state after the BS can be rephrased as one of the following two
orthogonal states,$|\phi_0>=\sqrt{T}|0>_a|H>_b+i \sqrt{R}|H>_a|0>_b$,
$|\phi_1>=\sqrt{T}|0>_a|V>_b+i \sqrt{R}|V>_a|0>_b.$
Here $R$ and $T$ are the reflectivity and transmissivity rate of the
BS, respectively. In our case $R$ and $T$ are 50\%/50\%. The $a$ and $b$
denote the path toward Alice's Faraday mirror, and
the path toward Bob site, respectively. We use $|0>_a$ and $|0>_b$ to represent
the vacuum state in the modes $a$ and $b$, respectively. As for the
events for which $D1$ detects a correct final polarization, the
initial states will collapse into two states of $|H>_a|0>_b$ or
$|V>_a|0>_b$. In fact, Bob will extract secure keys with Alice from
nondetection events.

In the experiment we use 1550nm DFB laser as signal source that is
actively modulated into about 1 ns pulse width. The intensity of
laser was attenuated to $0.1\sim 1$ photon/pulse on average at the
output of Alice's side. The total loss of the arm $b$ was measured
to be around 10.5 dB when photon goes through and returns back after
Bob's FM for case of an on-table test. The system works at a repetition rate of 100 KHz.
Currently this is mainly limited by the performance of feedback
loop. In addition, response time of the fiber switch of about
100 ns, and switching time of about 300 ns are needed to choose the
polarization of each pulse.

An active feedback control system is developed for phase stabilization.
Using a DWDM, we couple the signal light, and the 1570nm reference light
of a DFB CW laser, before their entering at the BS. When the system
works in feedback mode, Bob always selects to reflect the reference
light by choosing the optical switch to the FM. The reference light
goes through the interferometer via the same path of signal light,
and goes back through DWDM to a different output of optical
circulator. A highly sensitive APD-diode D0 is used to detect the intensity of the
output of interferometer. Thus, the reference light goes through the
same interferometer as the signal light, and experiences the same
path-difference. However the phase differences are different due to
different wavelengths of signal and reference light. Following a
similar analysis shown in \cite{Cho-Noh2009},
we can illustrate the phase difference with $n \triangle
L=\lambda_r(m_r+\varphi_r/2\pi)=\lambda_s(m_s+\varphi_s/2\pi)$. Here
$n$ is the refractive index, subscripts $r,s$ refer to reference and
signal light, respectively. $\triangle L$ is path difference. The
relative phase difference $\varphi_r,\varphi_s$ have a relation with
its wavelength $\lambda$ for a fixed setup. The $m$ is integer,
indicating phase difference being of integral multiple of its
wavelength. The intensity in the output of interferometer can be
given by $I=I_0 \cdot \cos^2(\frac{k \triangle L}{2})$ where $k$ is
wave number. By stabilizing the output intensity of reference to a
settled level, one can thus fix the path difference for signal light.
Experimentally, a homemade PI controller from cold-atom experiment
is modified and adjusted for experimental need. A voltage amplifier
is used to control the fiber stretcher (FS), which can fast change
the relative length of fiber in path $a$.  The PI controller
compares the output of APD-diode with a pre-set monitor level. If
the output voltage of APD-diode differs from a monitor level, PI
circuit will adjust the output voltage for compensating the change.
For our 1 km system, phase change is roughly in a level of
$0.1 \sim 0.5$ $rad/ms$. A total range of $20V (-10V \sim
+10V)$ output with subsequent amplification factor of $10\sim 40$ times enables
to make a 200$\pi$ rad adjustment through
fiber stretcher. A final interference visibility of better than 98\%
is observed for both the case of desktop test and the 1 km fiber cable
application, which enables successful operations of feedback
controlling system. Continuous running of about an hour for desktop
test, and 900 s for 1 km application have shown stable and reliable performances.

An FPGA electronics board is developed for controlling signal laser,
detectors and fiber switch. The timing is about 1$\mu s$ for signal
pulse generation and detection, and 9 $\mu s$ feedback time for
proportional-integral (PI) circuit, within a 10 $\mu s$ duty circle
for FPGA. We use one 1570 nm DFB laser as reference light.
We have firstly implemented the scheme in case of on-table test.
With PI feedback, a fringe visibility of better than 98\% could be
maintained. A raw key of 185651 bits is generated within a 3617s
running (amounting to 51.3 bits/s), with an average quantum bit
error rate (QBER) of 6.8\% when the average photon number is 0.5
photon/pulse. If we change the average photon number to be 0.05
photon/pulse, the raw key of 7891 bits will be produced with QBER of 4.2\%
on average. To test feasibility of CQC, we have extended the
scheme with a 1 km fiber cable connecting Alice and Bob's sites,
which will cause about 1 dB for channel loss. A fringe visibility of
about 98\% is managed to be fulfilled as well for 1 km fiber, which allows
generation of secure keys. Without any cover for
fibers on experimental desktop, a raw key of about 126 bits/s is
obtained with QBER of about 5.8\% for 900 s continuous
running, when the average photon number is chosen to be 1.0 photon/pulse. If the
average photon number is reduced to 0.5 photon/pulse, the measured
QBER is about 5.5\% while raw keys is decreased to be around 94
bits/s due to longer feedback time. See Table 1 for performance details.
We can see that for 1 km fiber realization, one can obtain considerable
raw key rates of 126 bits/s and good performance, with a quite low error rate of 0.6\% through $D3$
for channel monitoring, when using average photon number of 1.0.

In order to have a more comprehensive understanding for running of the experiment,
we present here a circumstantial analysis for experimental imperfections.
Consider we have two groups of data, one group that Alice and Bob choose the
same random numbers, and another group that they choose different random numbers.
The group with the same number will create a raw key if D1 is
detected with correct basis, which will be fulfilled if the
polarization is not disturbed. Another group will create a possible
wrong key if D1 is detected with correct basis, which will
contribute to an error. Here we take the case of 1 km experiment with
the mean photon number of 0.5 for an example to analyze. We attribute all the
errors coming mainly from three parts. The first part is from dark
counts and after-pulses of the detectors. The detectors we used have
a dark count possibility of each gate of about $10^{-5}$.
With 100 kHz working frequency, about 1 dark count is
created by detector, which will therefore contribute to about 0.7\% error rate,
as a total counts of $D1$ is about 70 counts per second in this case.
The after pulse will lead to about 0.5\% error rate due to a 1\%
probability of after pulse for detectors. The second part is from
finite extinct ratio of fiber switch. We have tested the static
extinction ratio of the fiber switch to be 20 dB, with a slight lower
of about 17 dB with regarding to the dynamic extinction ratio, which
amounts approximately to $1\%\sim 2\%$ error rate. The third one is coming
from imperfection of the optical alignment and the finite visibility
of feedback system, which will still lead clicks for $D1$ even if
Alice and Bob have different choices.
When Alice and Bob choose different random numbers, a visibility like
98\% is supposed to be achieved. This means that 1\% of the pulses are detected by
D1 and that the other 99\% are detected by D2.
The error rate can thus be estimated as $0.01/(0.01+0.25) = 3.8\%$, by
considering the fact that only 1/4 of the total pulses will be detected by D1 when
Alice and Bob choose same random number. All three parts therefore would
conduct to a QBER of around $6\%$ as confirmed by actual measured experimental data.

A preliminary security analysis is given in \cite{Noh2009}
by showing a new type of noncloning principle for orthogonal states: if
reduced density matrices of an available subsystem are nonorthogonal
and if the other subsystem is not allowed access, it is impossible
to distinguish two orthogonal quantum states $|\phi_0>$ and $|\phi_1>$.
without disturbing them. Moreover the scheme is robust to so-called
``photon-number splitting" attack. Since Eve cannot access
nondetection process that extracts secret keys\cite{Noh2009}, these
distinctive property provides a security advantage over existing schemes.
Consider in a typical noise channel case as analyzed in \cite{Guo2010}
in which possible upper bound for phase error rate could be less than QBER, our system
allows sufficiently to generate secure keys after one-way key
distillation process \cite{Shor-Preskill} for ideal case of source and detectors.

We have achieved the faithful and complete realization of counterfactual quantum communication, in which process
information carriers are seemingly not traveled in the quantum channel. From a desktop test to a setup with 1 km
fiber cable, we have given proof-in-principle demonstrations of CQC. This gives a confirming answer for feasibility
of CQC, and one can infer that the mere possibility for signal particles to be transmitted is sufficient to create
a secret key. We remark that, to ensure such possibility, partial signal particles still need to randomly travel
along quantum channel for detection of possible eavesdropping. One may wonder in the implemented scheme, signals
pass through the channel twice and thus suffer losses twice. It should be remarked that the CQC scheme is, in fact,
equivalent to the normal BB84 protocol for communication overhead of transmission loss, if one assists with a
quantum repeater through DLCZ scheme \cite{DLCZ} to teleport single photon signal directly between Alice and Bob.
Also one may see that the scheme requires to maintain long-term subwavelength stability of the path difference
between two arms of a long-distance interferometer, which can however be overcome by utilizing state-of-the-art
technology, similar to the one developed in \cite{Phase2007} for coherent optical phase transfer over 30 km fiber.
We can see that our implementation is based on currently available technologies, promising a novel conceptually
quantum communication system compared with existing systems. Presently the performance of our system is mainly
limited by speed of feedback. Consider our feedback system is only a modification of PI used in our cold-atom
experiments, significant improvements could be made by exploiting the method developed in \cite{Cho-Noh2009} for
interferometer stabilization, and by employing tailored design with a differential circuit, an auto-reset function
and exploiting adjacent channel in C band of 1550.12nm. Although technically challenging, our implementation in
conjunction with optimizing quality of interference via improved optical design, appears as a viable route toward
extendable realization up to long distance fiber.

\clearpage

\begin{figure}
\begin{center}
\epsfig{file=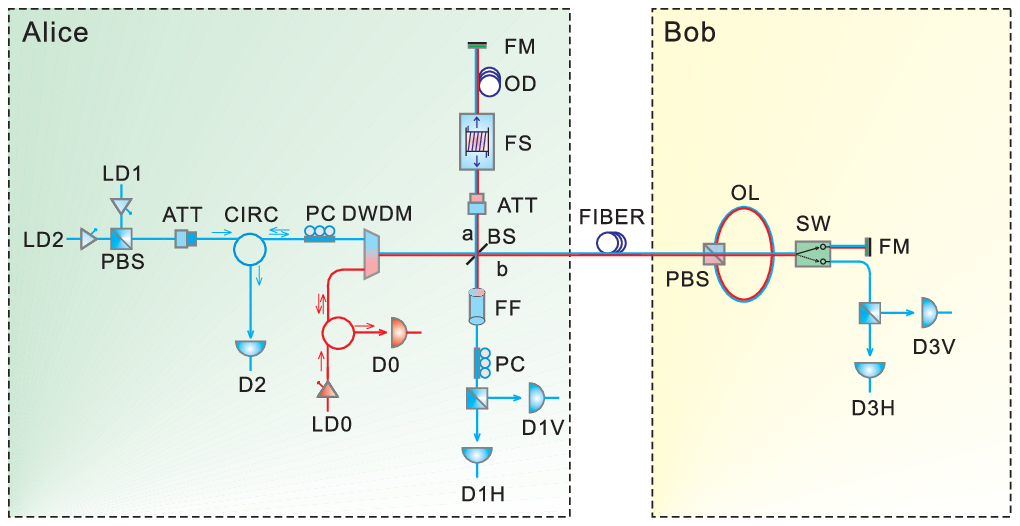,width=16cm}
\end{center}
\label{cf-setup}
\end{figure}
\noindent \textbf{Fig. 1.} Schematic setup of fiber based counterfactual
quantum cryptography system. A signal photonic pulse is sent from
Alice's side before passing through a optical circulator and then
splitting by a BS to travel along path $a$ and $b$. Alice chooses
randomly one of states of horizontal polarization $|H>$ representing
``0", and vertical polarization $|V>$ representing ``1". Depends on
Bob's choice coinciding with Alice's choice or not, the two split
pulses will be recombined in the BS and detected by detector $D2$
due to constructive interference, or a split pulse will go through
path $b$ and be blocked by detector $D3$. As a result of latter
case, interference is destroyed and $D1$ will detect a photon will
certain probability. Sifted keys will be established by selecting
only the events for which $D1$ alone detects a photon with a correct
final polarization state. Note that, when $D1$ clicks the photon
only travels inside Alice's secure zone, and there's no photon
travel to Bob's side, which excludes Eve's access for each signal
particle of the entire quantum system.
LD: laser diode, PBS: Polarization Beam Splitter, ATT: fiber attenuator, CIRC: optical
circulator, PC: polarization controller, BS: Beam
Splitter, FS: Fiber stretcher, DWDM: Dense wavelength division
multiplexer, FF: Fiber Filter, OD: Optical delay line, OL: Optical
delay, SW: fiber switch, FM: Faraday rotator mirror. D0,D1,D2,D3: detectors.
D1H, D1V are short notions for detector $D1$ of detecting horizontal
and vertical polarization, the same for D3H and D3V.

\clearpage
\begin{figure}
\begin{center}
\epsfig{file=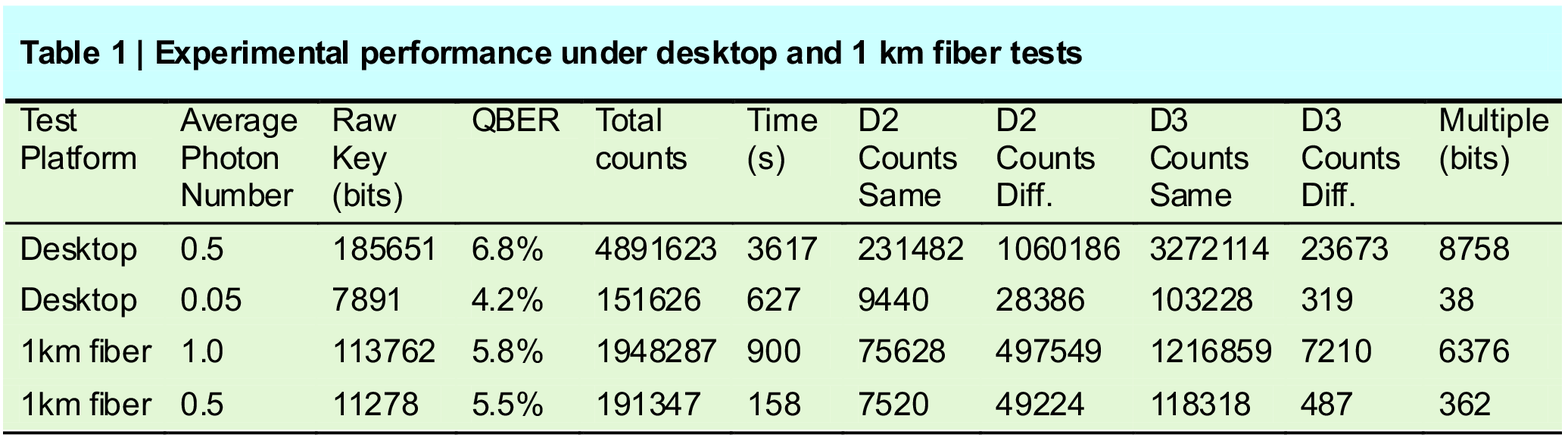,width=16cm}
\end{center}
\label{table}
\end{figure}
\noindent \textbf{Table 1.} Performance for counterfactual quantum cryptography system.
Table summarizing the various measured parameters characterizing
performances for desktop and 1 km fiber under different average
photon number. Besides the symbols introduced in the maintext, here we
use ``Total Counts" denoting sum of all the detectors $D1$,$D2$ and
$D3$. ``D2 Counts Same" and ``D2 Counts Diff." refer to counting of
$D2$ when Alice and Bob choose the same bit values and different
values, respectively. ``Multiple" means all the counts for events
that there are at lease two detectors clicks.


\begin{thebibliography}{99}
\bibitem{Noh2009}
T.G. Noh,
\textit{Phys. Rev. Lett.} \textbf{103,} 230501 (2009).

\bibitem{BB84}
C.H. Bennett, G. Brassard,
Quantum cryptography: public key distribution and coin tossing,
in \textit{Proceedings of the IEEE International Conferenceon Computers,
Systems and Signal Processing,} (Bangalore, India, 1984), pp. 175--179.

\bibitem{Lo-Chau1999}
H.-K. Lo, H.F. Chau,
\textit{Science} \textbf{283,} 2050 (1999).

\bibitem{Shor-Preskill}
P.W. Shor, J. Preskill,
\textit{Phys. Rev. Lett.} \textbf{85}, 441 (2000).

\bibitem{Gisin2002}
N. Gisin, G. Ribordy, W. Tittel, H. Zbinden,
\textit{Rev. Mod. Phys.} \textbf{74,} 145 (2002).

\bibitem{Hwang2003}
W.Y. Hwang,
\textit{Phys. Rev. Lett.} \textbf{91,} 057901 (2003).

\bibitem{GLLP}
D. Gottesman, H.-K. Lo, N. L\"utkenhaus, J. Preskill,
\textit{Quant. Inf. Comput.} \textbf{5,} 325 (2004).

\bibitem{Wang2005}
X.-B. Wang,
\textit{Phys. Rev. Lett.} \textbf{94,} 230503 (2005).

\bibitem{Lo-Ma-Chen2005}
H.-K. Lo, X.-F. Ma,  K. Chen,
\textit{Phys. Rev. Lett.} \textbf{94,} 230504 (2005).

\bibitem{Elliott2005}
C. Elliott \textit{et al.},
Current status of the DARPA Quantum Network, in
\textit{Quantum Information and Computation III,}
Donkor, E.J., Pirich, A.R., \& Brandt, H. E. eds.,
\textit{Proc. SPIE} \textbf{5815,} 138-149 (2005).

\bibitem{ILM}
H. Inamori, N. L\"utkenhaus, D. Mayers,
\textit{Eur. Phys. J. D} \textbf{41,} 599 (2007).

\bibitem{Pan-decoy}
C.-Z. Peng \textit{et al.},
\textit{Phys. Rev. Lett.} \textbf{98,} 010505 (2007).

\bibitem{America-decoy}
D. Rosenberg \textit{et al.},
\textit{Phys. Rev. Lett.} \textbf{98,} 010503 (2007).

\bibitem{EU-decoy}
T. Schmitt-Manderbach \textit{et al.},
\textit{Phys. Rev. Lett.} \textbf{98,} 010504 (2007).

\bibitem{Toshiba-decoy}
Z.-L. Yuan, A.W. Sharpe, A.J. Shields,
\textit{Appl. Phys. Lett.} \textbf{90,} 011118 (2007).

\bibitem{SECOQC}
M. Peev \textit{et al.},
\textit{New J. Phys.} \textbf{11,} 075001 (2009).

\bibitem{review2009}
V. Scarani, H.B. Pasquinucci, N.J. Cerf, M. Dusek, N. L\"utkenhaus, M. Peev,
\textit{Rev. Mod. Phys.} \textbf{81,} 1301(2009).

\bibitem{Qiang09}
Q. Zhang \textit{et al.},
\textit{New J. Phys.} \textbf{11,} 045010 (2009).

\bibitem{Guo2009}
W. Chen \textit{et al.},
\textit{IEEE Photonics. Tech. Letts.} \textbf{21,} 575 (2009).

\bibitem{Chen2010}
T.-Y. Chen \textit{et al.},
\textit{Opt. Exp.} \textbf{18,} 27217 (2010).

\bibitem{TokyoQKD}
M. Sasaki \textit{et al.},
\textit{Opt. Exp.} \textbf{19,} 10387 (2011).

\bibitem{ElitzurVaidman1993}
A.C. Elitzur, L. Vaidman,
\textit{Found. Phys.} \textbf{23}, 987 (1993).

\bibitem{Mitchison-Jozsa2001}
G. Mitchison, R. Jozsa,
\textit{Proc. R. Soc. Lond. A} \textbf{457} 1175 (2001).

\bibitem{Kwiat1999}
P. G. Kwiat \textit{et al.},
\textit{Phys. Rev. Lett.} \textbf{83,} 725 (1999).

\bibitem{kwiat2006}
O. Hosten, M.T. Rakher, J.T. Barreiro, N.A. Peters, P.G. Kwiat,
\textit{Nature} \textbf{439,} 949 (2006).

\bibitem{noncommuting2007}
V. Parigi, A. Zavatta, M.S. Kim, M. Bellini, \textit{Science} \textbf{317,} 1890 (2007).

\bibitem{Guo2010}
Z.-Q. Yin, H.-W. Li, W. Chen, Z.-F. Han, G.-C. Guo,
\textit{Phys. Rev. A} \textbf{82,} 042335 (2010).

\bibitem{Zeng2011}
M. Ren, G. Wu, E. Wu, H. Zeng,
\textit{Laser Phys.} \textbf{21}, 755-760 (2011).

\bibitem{DLCZ}
L.-M. Duan, M.D. Lukin, J.I. Cirac, P. Zoller,
\textit{Nature} \textbf{414}, 413 (2001).

\bibitem{Phase2007}
S.M. Foreman, A.D. Ludlow, M.H.G. de Miranda, J.E. Stalnaker, S.A. Diddams, J. Ye,
\textit{Phys. Rev. Lett.} \textbf{99,} 153601 (2007).

\bibitem{Cho-Noh2009}
S.-B. Cho, T.-G. Noh,
\textit{Opt. Exp.} \textbf{17,} 19027 (2009).

\bibitem{Acknowledgements}
We acknowledge the financial support from the CAS, the National Fundamental
Research Program of China under Grant No.2011CB921300, the National High
Technology Research and Development Program (863 Program) of China
under Grant No.2009AA01A349, the NNSFC and the Fundamental Research Funds
for the Central Universities. The authors are grateful for valuable discussions with
Dr. Xian-Min Jin.

\end{thebibliography}
\end{document}